\documentclass[twocolumn,10 pt,showpacs,preprintnumbers,amsmath,amssymb]{revtex4}
\usepackage{amssymb}
\usepackage{graphicx}
\begin{document}

\title{Magnetoresistance, noise properties and the Koshino-Taylor effect in the quasi-1D oxide $KRu_4O_8$}
\author{Alain Pautrat}
\affiliation{Laboratoire CRISMAT, UMR 6508 du CNRS, ENSICAEN et Universit\'{e} de Caen, 6 Bd Mar\'{e}chal Juin, F-14050 Caen 4.}
\author{Wataru Kobayashi}
\affiliation{Graduate School of Pure and Applied Sciences, University of Tsukuba, Ibaraki 305-8571, Japan; Tsukuba Research Center for Interdisciplinary Materials Science (TIMS), University of Tsukuba, Ibaraki 305-8571, Japan;
PRESTO, Japan Science and Technology Agency, Saitama 332-0012, Japan.}

\begin{abstract} 
The low temperature electronic and galvanomagnetic transport properties of the low dimensional oxide $KRu_4O_8$ are experimentally considered.
A quadratic temperature variation of the resistivity is observed to be proportional to the residual resistivity. It shows the role of inelastic 
electron scattering against impurities, i.e. a large Koshino-Taylor effect, rather than a consequence of strong electronic correlations.
In the same temperature range, the Kohler rule is not fulfilled. The resistance noise increases also sharply, possibly due to a strong coupling
 of carriers with lattice fluctuations in this low dimensional compound.
\end{abstract}

\maketitle
\section{Introduction}

Hollandite crystals have been the subject of a number of investigations, due to their interesting physical
 properties intimately related to their low dimensional character \cite{foo,mao,wata1,wata2,alain,maignan}.
 The hollandite structure is based on double chains of $MO_6$ octahedra ($M$=transition element such as Ru, Rh...) conferring a quasi-one dimensional (1D) structure.
 Recent calculations of the electronic structure of $KRu_4O_8$ indicate that the Fermi surface itself has a quasi-1D nature,
 with two sheet-like surfaces which are parallel and separated by a wave vector $\pi /c$ \cite{toriyama}. This result is in agreement with reports of
 anisotropic electronic conduction mechanisms and possible dimensional cross-over in hollandites \cite{wata1,alain}.

 In a pure 1D electronic interacting system, only collective excitations can exist. A Fermi liquid description is not possible
 and is replaced by a Tomonaga-Luttinger liquid with spins and charges separation\cite{giam, voigt, schultz}. 
Recently, a strong violation of the Wiedeman-Franz law
 in the low dimensional purple bronze $Li_{0.9}Mo_6O_{17}$ was found consistent with Tomonaga-Luttinger liquid theory \cite{hussey1}.
  In a low dimensional system, it is well known that a charge density wave instability due the electron-lattice coupling can also be expected,
 resulting in a low temperature resistivity upturn.
This is not observed in $KRu_4O_8$ which shows a metallic behavior along the conducting chains with no anomaly down to low temperatures (T = 1.8 K).
 This is in agreement with the imperfect
 nesting of the Fermi surface reported in \cite{toriyama}.
The strong warping of the two parallel pairs of Fermi surfaces of $KRu_4O_8$ \cite{toriyama} 
indicates also that the system is not perfectly 1D and that interchains coupling is substantial.
 As a consequence, there is a place for physical properties
 at a dimension larger than one at low temperature, such as Fermi liquid transport properties  \cite{giam, voigt, schultz}.
 For example, 3D Fermi liquid properties are
recovered at low temperature in the Quasi-1D cuprate $PrBa_2Cu_4O_8$ \cite{hussey2}.
 Opposite temperature dependence for the in plane and out of plane resistivity have been reported in $KRu_4O_8$ for a temperature T$>$180 K \cite{wata2}, 
as expected for quasi-1D electronic transport showing incoherent transverse conduction.
 However, the electronic anisotropy appears strongest at low temperature, a result which is not expected in quasi-1D
 materiels where the interchains hopping takes place at low temperature leading to a more 3D behavior.
Note that another hollandite $Ba_{1.2}Rh_8O_{16}$ \cite{alain} shows a behavior more in line with this expected dimensional crossover. 
 It suggests that if $KRu_4O_8$ presents the quasi-1D Fermi surface, some additional resistive processes could take
 place at low temperature.

The characterization of a Fermi liquid is often based upon the thermal variation of the resistivity. 
A quadratic term $AT^2$ is expected at low temperature for Landau
 quasiparticle-quasiparticle interaction, i.e. a Fermi liquid behavior, for dimensions greater or equal to two \cite{giam}.
 Its observation can indicate that the system has free quasiparticle excitations and invalidates \textit{a priori}
 the existence of a more exotic electronic state imposed by a 1D motion of interacting electrons \cite{giam}.
 Then, the transition from a $T^2$ to a different power law can indicate a transition from a Fermi Liquid to non Fermi liquid behavior,
 i.e. from a 2D or 3D to a 1D behavior. 
 Experimentally, the analysis of the low temperature resistivity is however not trivial due to the
 unavoidable sample disorder and impurities \cite{capogna}. In particular, metallic oxides present different kind of structural defects and
 disorder (cationic and oxygen disorder for instance) which result in a large residual resistivity $\rho_0$. 
 
 Assuming Matthiesen rule to hold, the resistivity can be written in the form

\begin{equation}
\rho=\rho_0+AT^2+\rho_{ph}
\end{equation}

with $\rho_{ph}$ the resistivity due to phonon scattering.
At a low temperature, $\rho_{ph}$ is negligible and the $T^2$ dependence can be observed.
As discussed in \cite{garba1}, the quadratic temperature dependence can also arise from different mechanisms:
 not only carrier-carrier scattering through strong coulomb interactions \cite{giam, barber} but also
from carrier magnetic moment scattering \cite{magnet}. Another possible mechanism is the so-called Koshino-Taylor effect
describing inelastic scattering against impurities \cite{koshino, taylor}. This effect is particular in the sense
 that $\rho_0$ can not be simply subtracted to extract the intrinsic resistivity.
Including all terms, the low temperature resistivity should be rewritten as \cite{garba1} 
 
\begin{equation}
\begin{split}
\rho=\rho_0(1+B T^2)+A T^2+\rho_{ph}\\
=\rho_0+(B \rho_0+A) T^2+\rho_{ph}\\
=\rho_0+\alpha T^2+\rho_{ph}
\end{split}
\end{equation}
  
 In principle, extracting the coefficient $A$ of strong correlations requires then to plot the slope $\alpha$ of $\rho$ versus $T^2$
 (at low temperature where the phonon contribution to the resistivity freezes out). Different values
 of $\rho_0$ can be obtained by tuning the pressure or by measuring samples with different amount of disorder.
 If $\alpha$ does not depend on $\rho_0$, then a Fermi-liquid scenario can be inferred. Otherwise, inelastic scattering by impurities
 is likely the relevant mechanism.
In classical metals, the Koshino Taylor effect is a small effect but can be observed \cite{mahan}. It has been shown to
 be more important for at least two recent and highlighted materials, in superconducting
 pnictides \cite{garba2} and in graphene \cite{cano}. In the latter case, the low dimensional character of graphene was shown to be a possible reason
 of a reinforced strength of the Koshino-Taylor effect compared to ordinary metals.

We report here measurements of the magneto-transport properties of the metallic and paramagnetic low dimensional oxide $KRu_4O_8$ for samples
 from the same batch but with different residual resistivities.
 The Koshino-Taylor effect is observed to be strong and
 dominant at low temperature, and is simultaneous with a large resistance noise.

\section{Experimental}
$KRu_4O_8$ is an alkaline ruthenium hollandite. Single crystals of $KRu_4O_8$ were grown by using a flux method as described in
\cite{wata2}. The obtained crystals have
a needle-like form with a typical dimension of 1 mm $\times$ 0.05 mm $\times$ 0.05 mm. Unit-cell
parameters are a=9.913(5) \AA, and c=3.108(5) \AA.
For the resistance measurements, four gold wires with a diameter of 20 $\mu$m were glued with silver paste. To obtain a low contact resistance, the
sample was annealed at 673 K during 10 min in air. The samples present low residual resistivities,
 typically in the $\mu \Omega.cm$ range, close to the best crystals of $Sr_2RuO_4$ where unconventional superconductivity has been observed \cite{srruo}. 
Transport measurements have been performed in a PPMS Quantum Design equipped
 with a 14T magnet and a rotator. For the noise measurements, we use a home made sample holder.
 The sample is biased with a noise free current supply of 100 mA, voltage time series are amplified
 using ultra low noise preamplifiers (SA-400F3) enclosed in a thick box and the signal is then anti aliased
 and Fourier transformed in real time using a PCI acquisition card (see details in \cite{scola}). In what follows, 
$\rho$ will stand for the resistivity along the chains, i.e. along the c-axis. 
 
\section{Galvanomagnetic properties and Koshino-Taylor effect}

 In the figure 1 is shown a typical $\rho(T)$ curve, very similar as previous reports \cite{foo, wata2}.
 The metallic character is pronounced, and the residual resistivity is low for an oxide (in the $\mu\Omega$.cm range),
 giving a RRR ratio $\rho_{300 K}/\rho_{2K}\approx 96$ for the purest sample.
Contrarily to another quasi-1D hollandite $Ba_{1.2}Rh_{8}O_{16}$ \cite{alain}, no upturn of the zero field resistivity 
is observed at low temperature. 
An estimation of the electronic mean free path $\ell $ along the chain using the relation $\rho=a^2 \pi \hbar /(2 e^2)\ell$ \cite{hussey} 
gives a rather large mean free path $\ell\approx$ 130 nm at 2K.

At low temperature where the phonon resistivity freezes out, a classical approach is to fit $\rho(T)$ as a power
 law $T^{\beta}$. $\beta$=2 is expected for electron-electron (Baber-Landau) scattering  \cite{barber}, carrier-magnetic moment scattering \cite{magnet} and scattering by impurities and dislocations \cite{garba1, koshino, taylor, fisher}.
 Since $KRu_4O_8$ is a Pauli paramagnet \cite{foo, wata2}, only the first and third cases have to be considered.
 An exponent of $\beta$=2.7 has been previously reported in $KRu_4O_8$ for temperature lower than 62 K and was discussed as 
arising from electron-electron umklapp scattering in a 1D conductor \cite{wata2}, as described by Oshiyama et al \cite{umklapp}. 
However, the two pairs of warped Fermi surface needed to realize this umklapp scattering were not observed in \cite{toriyama}.    
We have measured three samples of the same batch exhibiting different residual resistivity $\rho_0$, implying different impurities concentration. 
Depending on $\rho_0$, the thermal variation of the resistivity at the lowest temperatures (typically below 50 K) exhibits different power
 law dependences. This variation is at least partly due to the resistivity saturation, i.e. its intrinsic nature is difficult to prove.
Contrarily, a $T^2$ component is observed for all samples and for temperatures slighly larger than the temperature where $\rho$ starts to saturate (inset of fig.1).
An important observation is that the slope $\alpha$ of $\rho=f(T^2)$ in the equation(2) is different for the three measured samples (fig.2),
 with differences significantly larger than the uncertainty in the values of $\alpha$ (error bars are due to approximations when measuring the sample size and the distance between contacts).
 As discussed above, it implies that electron-electron scattering alone can
 not explain the $T^2$ dependence of the resistivity. 
As shown in fig.2, the data can be describe by the equation $\alpha=A+B\rho_0$, with a slope $B\approx 5 \times 10^{-4} K^{-2}$ and $A\approx 0$.
We conclude that the Fermi liquid component $A$ is extremely low here, and the Koshino-Taylor component is dominating. Note that the Sommerfeld coefficient 
of $KRu_4O_8$ is about 3 $mJ/mol_ {Ru} K^2$ \cite{foo}. It corresponds to a moderately renormalized mass of carriers,
 in agreement with a "non strongly correlated" character implied by the very low value of $A$.
The Value of $B$ is orders of magnitude larger than Koshino-Taylor effects reported in the literature.
 More precisely, Taylor's prediction was $B\sim 0.1 / \theta_D^2$ for conventional metals, i.e  $B\sim 7 \times 10^{-8} K^{-2}$ using $\theta_D \approx 370 K$ \cite{foo}
 what is largely lower than our experimental value. 
 In the layered iron arsenide $LaFeAsO_{0.9}F_{0.1}$, a value of  $B\approx 6 \times 10^{-6} K^{-2}$ was reported \cite{garba2}, and $B\approx 10^{-5} K^{-2}$ in Nb-Ti \cite{garba1}.
 Our value is closer to what was reported in $Al_3Zr$ where $B\approx 1 \times 10^{-4} K^{-2}$ \cite{fisher}. An unusual large transverse electron-phonon coupling
 was then suggested \cite{fisher}.
A large Koshino-Taylor effect was recently discussed for the graphene, taking into account the low dimensional character of the lattice fluctuations \cite{cano}. 
 We propose that a similar argument should be applied to $KRu_4O_8$ which presents a quasi-1D lattice. The predicted $TlnT$ dependence of the Koshino-Taylor correction to the resistivity \cite{cano} is not observed however,  
in a reasonable temperature range. Note that a $T^2lnT$ dependence gives a sightly better agreement than a pure $T^2$ dependence in our data because it extends down to lowest temperature.
This dependence was deduced in the extension of electron phonon impurity interference theory of Reizer-Sergueev at low dimension \cite{reizer}.
 It could explain the non Fermi liquid like dependence reported in some $KRu_4O_8$ samples \cite{wata2}. We would like to stress that if it is difficult to conclude on
the exact temperature dependence, i.e. $T^2$ or $T^2lnT$, the dependence on the residual resistivity
is a strong support to consider here inelastic scattering against impurities as a major process.

The magnetoresistance MR=$(\rho (B)- \rho (0))/\rho (0)$ is positive for all temperatures and shows strong angular angular dependence (see fig.3),
 being consistent with both a Lorentz force driven MR and a large anisotropy of the Fermi surface
 in our geometry. A MR of more than 90 \% is observed at 14T and at low temperature (T=1.8 K, fig.3).
This large value is consistent with the large mean free path of $KRu_4O_8$. 
As shown in the figure 3, the MR at low field has a non quadratic field dependence, and can be fitted by a 1.3 power law.
A conventional and metallic MR is supposed have a low field $B^2$ component, but lowest exponents in the range 1.3-1.5 have been already reported
 in low dimensional bronze oxides \cite{bronze}, organic conductors \cite{kriza} or in metallic nanowires such as Bi \cite{bi}.
 No clear saturation of the MR can be observed up to 14T. At the lowest temperature that we have measured (T=1.8K), the MR tends to curve downward at large field,
 indicating an incipient saturation. This can be described by a two band model with uncompensated carriers \cite{twoband}, similarly to \cite{bronze}. 
 
The Kohler plot $\Delta  \rho/\rho_0 = F(B/T)$ of MR shows that the galvanomagnetic processes
can be separated in two temperature ranges (fig.4). For $T<T^*\approx $18 K, the MR curves are superimposed on the same plot.
For $T>T^*$, the Kohler rule is not supported by the data. As a general rule, the Kohler rule applies in the case of a single scattering process \cite{kohler}.
This is the case for $T<T^*$ where the resistivity becomes temperature independent and tends to the residual resistivity value.
 At larger temperature $T>T^*$, the contribution of the Koshino Taylor resistivity becomes apparent and breaks the Kohler rule, as expected if the two scaterring processes have different magnetic field dependence.
 Note that the magnetic field dependence of MR is unchanged when crossing $T^*$, strongly suggesting that the Koshino-Taylor mechanism contributes to the zero field resistivity but not significantly to the MR.


\section{Noise properties}
Noise is a powerful tool to investigate the low frequency dynamics of electronic processes \cite{mike}, and appears
 particularly interesting for low dimensional systems since fluctuations increase in principle when dimensions shrink.
 The experimental quantity is here the noise spectrum $S_{VV}$ of the fluctuating voltage from the sample, using the set-up described in the experimental part.
The noise values present a quadratic dependence with the applied current, as expected for noise driven by the fluctuations of resistance.
 After analyzing the time series of noise spectra ,
there is no evidence of non Gaussian components of the noise (it corresponds to a white so called second spectrum \cite{second}).
It can be assumed then that the noise arises from uncorrelated random processes.
 
 We have measured $S_{VV}$ of $KRu_4O_8$ as function of the temperature under a constant 100 mA current.
 Noise values are large when comparing with conventional metals, but also large compared to another Hollandite
 $Ba_{1.2}Rh_{8}O_{16}$ where the noise was not measurable 
with the same experimental set-up \cite{alain}.
To discuss normalized values between different materials of different resistances, the Hooge parameter $\gamma= S_{VV}/V^2 n_c \tau f^{\alpha}$ is usually employed \cite{hooge}.
$n_c$ is the carrier density ($n_c\approx 2.10^{22}$ cm$^{-3}$ \cite{wata2}), $\tau$ is the noisy (probed) volume and $\alpha\approx$ 2 here. We deduce $\gamma\sim 450$, i.e. four order of magnitude larger than metallic values which are around 0.01 \cite{hooge}.
Such a large value is not rare in oxides, as for example reported in colossal magnetoresistance manganites, $Fe_3O_4$ ($\gamma\sim 80$) and in $CrO_2$ ($\gamma\sim 2000$) \cite{raquet}.
 Note that for low dimensional conducting processes, assuming that the noisy volume $V$ corresponds to the whole volume between
the probes is an overestimation if inhomogeneous charge propagation paths exist. 
The temperature dependence of $\gamma$ is shown in fig. 5.
Interestingly, it increases sharply in the same temperature range where the $T^2$ component of resistivity dominates, and then tends to saturate at low temperature.
It shows a direct correlation between the change in the scattering processes and the electronic fluctuations. Consequently, we have to associate
 the reinforced resistance fluctuations with the scattering from impurities due to the Koshino-Taylor effect.
 It has to be emphasized that the spectral shape of the noise is $1/f^{\alpha}$ with $\alpha\approx 2$ for the whole temperature range (inset of fig.5).
  Noise usually arises from the activated motion of defects, dislocations or impurities.
These processes can have an intrinsic Lorentzian spectrum which presents a $1/f^2$ dependence for frequencies large enough \cite{mike}.
 However, at a macroscopic scale,
the distribution of energy barriers results in a $1/f$-like noise \cite{ddh}. $\alpha\approx 2$ is then rarely observed in bulk materials
compared to $\alpha\approx 1$ \cite{mike}.
For example $1/f^2$ noise can be caused by stress induced dislocations or strain relaxations \cite{bertotti},
 but it has been rarely observed in metals where $1/f$ noise dominates \cite{giord} except when the sample was submitted
 to very large strains \cite{nat}. A $1/f^2$ power spectrum corresponds also to a Brownian noise and arises directly from the statistics of a 1D random walk.
 We propose that the $1/f^2$ dependence of the spectrum is here characteristic of the 1D nature of charges fluctuations. Of course, it would be necessary to perform
 similar experiments in other 1D or quasi 1D oxides to strengthen this result.
 The strong increase of the normalized noise for T$<$150 K appears when the $T^2$ component of resistivity appears, i.e. the large Koshino-Taylor effect, 
revealing a strong coupling between the charges fluctuations and the fluctuations of the lattice. We note that
 an anomalous temperature dependence of the Seebeck coefficient was reported in the same temperature range \cite{wata2}.

 In conclusion, we have measured the galvanomagnetic properties and the resistance noise of quasi-1D hollandites $KRu_4O_8$.
 Despite the relative cleanest of the samples revealed by the low residual resistivity, a very
 large inelastic scattering of carriers by impurities is revealed by the analysis of the transport properties. It can be due to the importance of the local fluctuations of the lattice
 in this highly anisotropic material as discussed in \cite{cano}. Accordingly, the resistance noise is strongly reinforced at
 low temperature and the $1/f^2$ spectral form of the noise is likely characteristics of the 1D nature of the carriers paths.

\newpage
\begin{figure}[t!]
\begin{center}
\includegraphics*[width=8.0cm]{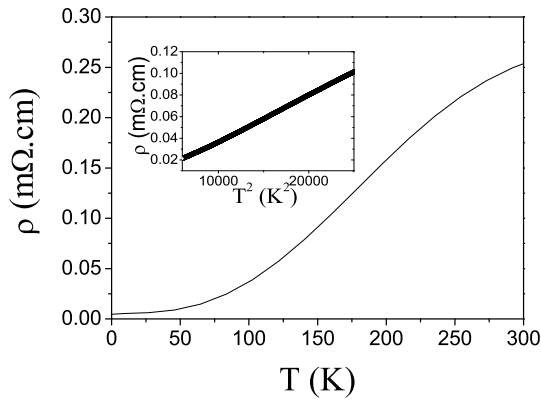}
\end{center}
\caption{Resistivity of $KRu_4O_8$ as function of the temperature (I=500 $\mu$A).
Inset: Resistivity of $KRu_4O_8$ as function of T$^2$.}
\label{fig.1}
\end{figure}

\newpage
\begin{figure}[t!]
\begin{center}
\includegraphics*[width=8.0cm]{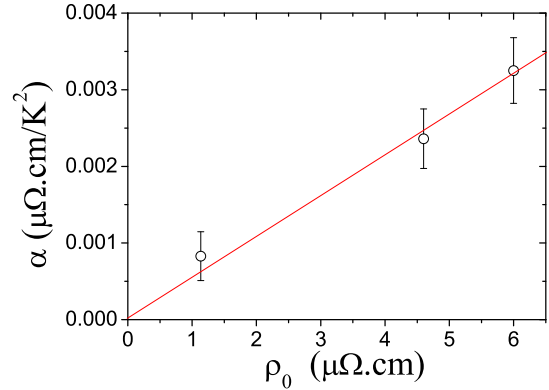}
\end{center}
\caption{Slope of $\rho$=$f(T^2)$ for different values of the residual resistivity $\rho_0$, each corresponding to a different crystal from the same batch.
 The dotted line is a linear fit discussed in the text and expected for the Koshino-Taylor mechanism.}
\label{fig.2}
\end{figure}

\newpage
\begin{figure}[t!]
\begin{center}
\includegraphics*[width=8.0cm]{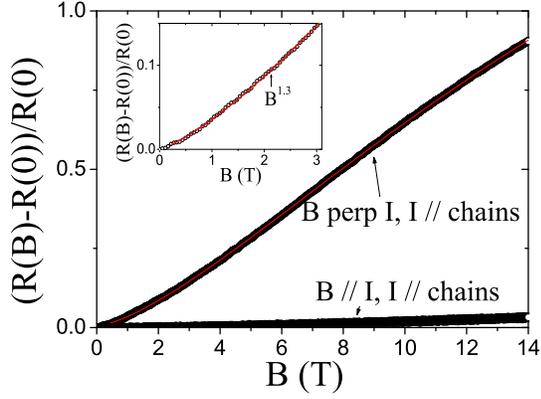}
\end{center}
\caption{Magnetoresistance $MR= (\rho(B)-\rho(0))/\rho(0)$ as function of the magnetic field for different geometries and for T=1.8 K. 
The solid line is a fit using a two bands model \cite{twoband}. In the inset is shown the low field part to evidence the B$^{1.3}$ dependence.}
\label{fig.3}
\end{figure}

\newpage
\begin{figure}[t!]
\begin{center}
\includegraphics*[width=8.0cm]{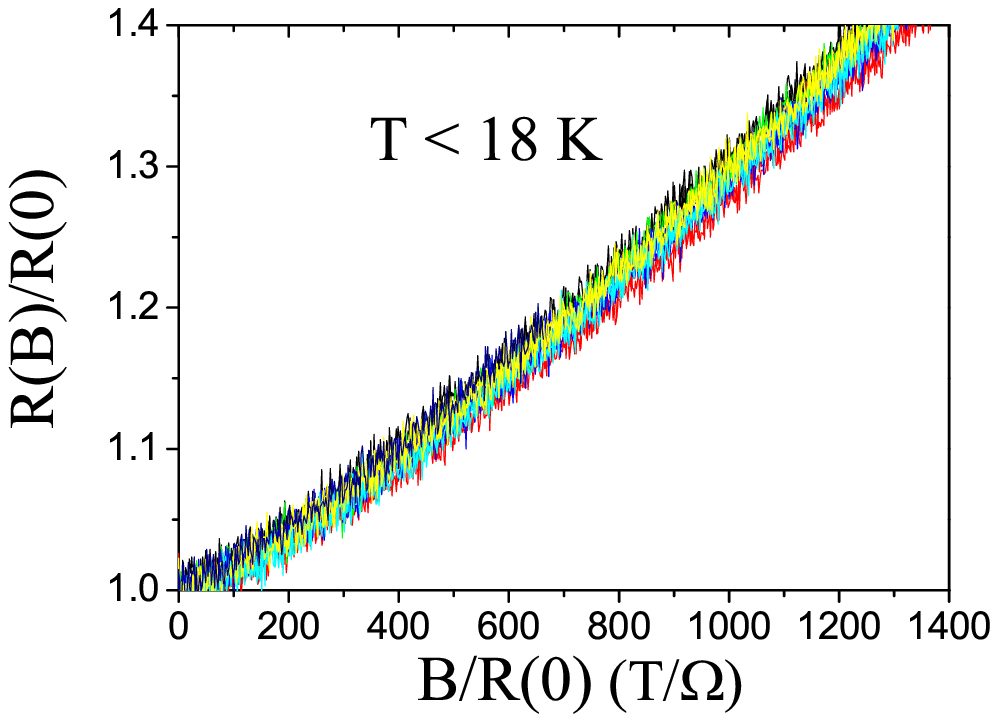}
\includegraphics*[width=8.0cm]{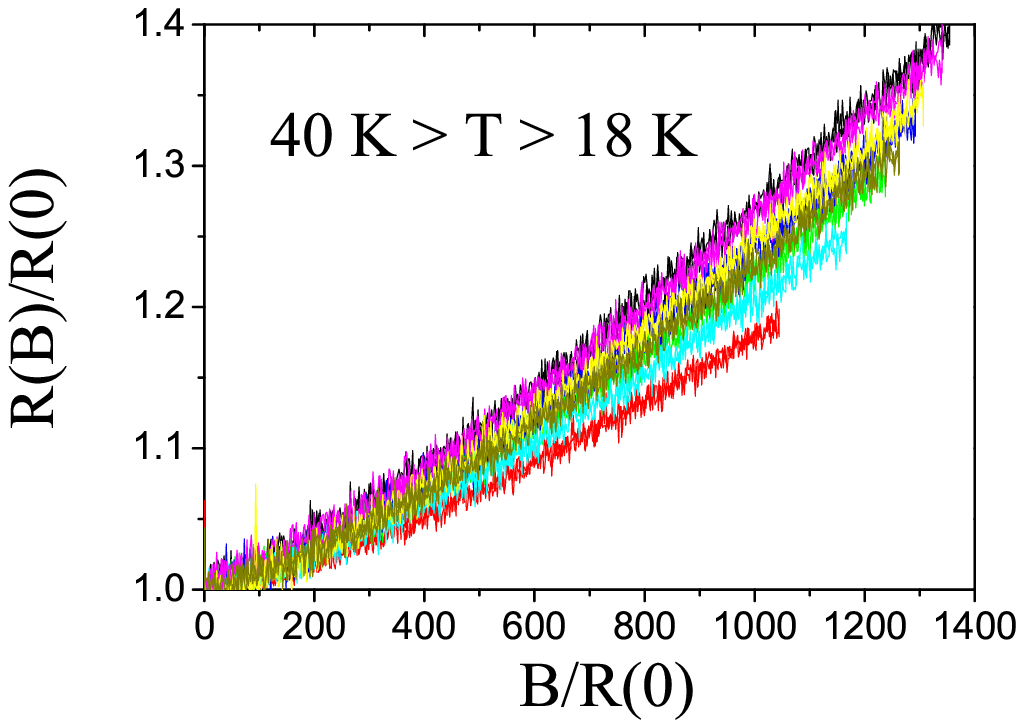}
\end{center}
\caption{Kohler Plot of the magnetoresistance for $T < 18K$ (top) and $T>18 K$ (bottom).
 The Kohler rule is fulfilled at low temperature but not when the resistivity starts to be temperature dependent.}
\label{fig.4}
\end{figure}


\newpage
\begin{figure}[t!]
\begin{center}
\includegraphics*[width=8.0cm]{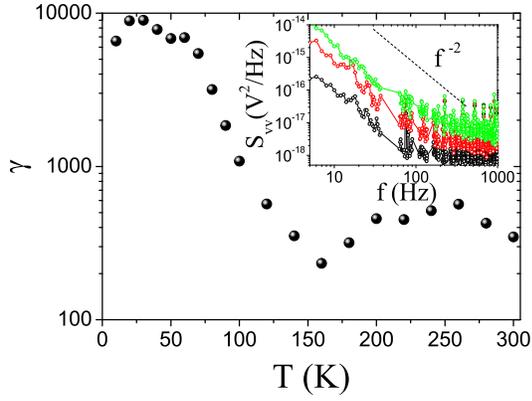}
\end{center}
\caption{Hooge parameter $\gamma$ (normalized noise value) as function of the temperature. Note the strong increase of $\gamma$ for $T<150K$.
 In the inset are shown the typical voltage noise spectra at $T=$ 300, 200, 100 K from top to bottom ($I=$ 100 mA).}
\label{fig.5}
\end{figure}

\end{document}